\documentclass[11pt]{article}

\usepackage[margin=1in]{geometry}

\usepackage{natbib}

\usepackage[utf8]{inputenc}
\usepackage{amsmath, amssymb, amsthm}

\usepackage{hyperref}
\usepackage{xcolor} 

\hypersetup{
    colorlinks=true,
    linkcolor=red,
    citecolor=blue,
    filecolor=blue,
    urlcolor=magenta,
    linktocpage=true
}

\usepackage{subcaption}
\usepackage{graphicx}
\graphicspath{{figures/}}
\usepackage{float}

\usepackage[linesnumbered,ruled,vlined]{algorithm2e}

\SetCommentSty{mycommfont}
\SetKwInput{KwInput}{Input}                
\SetKwInput{KwOutput}{Output}              

\usepackage{listings} 

\usepackage[newfloat, frozencache=true]{minted}

\usepackage{caption}
\newenvironment{code}{\captionsetup{type=listing}}{}
\SetupFloatingEnvironment{listing}{name=Code example}

\usepackage{cprotect}

\usepackage{makecell}

\usepackage{xcolor} 
\definecolor{YellowOrange}{RGB}{255, 153, 20}
\definecolor{MaxBlueGreen}{RGB}{8, 189, 189}

\newcommand{\spa}[1]{^{(#1)}}

\newtheorem{proposition}{Proposition}[section]

\newtheorem{remark}{Remark}[section]
\newtheorem{definition}{Definition}[section]

  \newcommand{\coding}[1]{{\tt #1}}

\newcommand{\loss}{\ell} 
\newcommand{\pen}{p} 

\newcommand{\coef}{\beta} 
\newcommand{\inter}{\coef_0} 

\newcommand{\estCoef}[1]{\widehat{\coef}\spa{#1}}  
\newcommand{\estInter}[1]{\widehat{\inter}\spa{#1}}  

\newcommand{\initCoef}{\estCoef{0}} 

\newcommand{\tuneparam}{\lambda} 

\newcommand{\sigmoid}{\text{sig}} 
\newcommand{\soft}{\text{soft}} 
\newcommand{\huber}{\text{huber}} 
\newcommand{\knot}{\text{knot}} 
\newcommand{\til}{\text{tilted-abs}} 

\newcommand{\tik}{\text{T}} 
\newcommand{\gen}{\text{M}} 

\newcommand{\enet}{\alpha} 

\newcommand{\sampWeight}{s} 
\newcommand{\offset}{t} 

\newcommand{\groups}{{\tt groups}}  

\newcommand{\adptExp}{\gamma} 

\newcommand{\support}{{\tt support}}  

\newcommand{\trans}{t} 

\newcommand{\prox}{\textsc{prox}}  

\newcommand{\lr}{\alpha}  

\newcommand{\interSolAtZero}{\inter^{null}}

\newcommand{\nuc}[1]{||#1||_{*} } 

\newcommand{\killerLbd}{\tuneparam_{\text{killer lbd}}} 

\newlength\dunder
\usepackage{xspace} 

\newcommand{\Pyaglm}{\href{https://github.com/yaglm/yaglm}{yaglm}\xspace}

\newcommand{\Psklearn}{\href{https://scikit-learn.org/stable/}{scikit-learn}\xspace}
\newcommand{\Pglmnet}{\href{https://glmnet.stanford.edu/articles/glmnet.html}{glmnet}\xspace} 

\newcommand{\Plightning}{\href{https://github.com/scikit-learn-contrib/lightning}{lightning}\xspace} 

\newcommand{\Ppicasso}{\href{https://github.com/jasonge27/picasso}{picasso}\xspace} 
\newcommand{\Pceler}{\href{https://github.com/mathurinm/celer}{celer}\xspace} 
\newcommand{\Pandy}{\href{https://github.com/mathurinm/andersoncd}{andersoncd}\xspace} 
\newcommand{\Pstatsmodels}{ \href{https://www.statsmodels.org/}{statsmodels}\xspace}

\newcommand{\Ppyglmnet}{\href{https://github.com/glm-tools/pyglmnet}{pyglmnet}\xspace}

\newcommand{\Pgrpreg}{\href{https://github.com/pbreheny/grpreg}{grpreg}\xspace} 
\newcommand{\Pncreg}{\href{https://cran.r-project.org/web/packages/ncvreg/}{ncreg}\xspace} 
\newcommand{\Pgenlasso}{\href{https://cran.r-project.org/web/packages/genlasso/index.html}{genlasso}\xspace} 
\newcommand{\Plassopack}{\href{https://statalasso.github.io/docs/lassopack/}{lassopack}\xspace} 

\newcommand{\Pcvxpy}{\href{https://www.cvxpy.org/}{cvxpy}\xspace}

\newcommand{\Ppyunlocbox}{\href{https://github.com/glm-tools/pyglmnet}{pyunlocbox}\xspace}

\newcommand{\PGGMncv}{\href{https://github.com/donaldRwilliams/GGMncv}{GGMncv}\xspace}

\title{yaglm: a Python package for fitting and tuning generalized linear models that supports structured, adaptive and non-convex penalties}

\author{Iain Carmichael\footnote{idc9@uw.edu}, Thomas Keefe, Naomi Giertych, Jonathan P Williams}
\date{\today\footnote{This is a living draft that will evolve until the package development stabilizes.
We welcome contributors to the package who will be added as coauthors to the manuscript.
The code lives at \url{https://github.com/yaglm/yaglm}.
 }}

\begin{document}

\maketitle

\begin{abstract}
The \Pyaglm package aims to make the broader ecosystem of modern generalized linear models accessible to data analysts and researchers.
This ecosystem encompasses a range of loss functions (e.g. linear, logistic, quantile regression), constraints (e.g. positive, isotonic) and penalties.
Beyond the basic lasso/ridge, the package supports structured penalties such as the nuclear norm as well as the group, exclusive, fused, and generalized lasso.
It also supports more accurate adaptive and non-convex (e.g. SCAD) versions of these penalties that often come with strong statistical guarantees at limited additional computational expense.
\Pyaglm comes with a variety of tuning parameter selection methods including: cross-validation, information criteria that have favorable model selection properties, and degrees of freedom estimators.
While several solvers are built in (e.g. FISTA), a key design choice allows users to employ their favorite state of the art optimization algorithms.
Designed to be user friendly, the package automatically creates tuning parameter grids, supports tuning with fast path algorithms along with parallelization, and follows a unified scikit-learn compatible API.
\end{abstract}


\tableofcontents{}

\section{Introduction}

Penalized \textit{generalized linear models} (GLMs) and other supervised M-estimators\footnote{Quantile and huber regression are examples of supervised M-estimators.} are some of the main tools used in modern statistics \citep{hastie2019statistical, fan2020statistical}. 
Given covariates $\{x_i\}_{i=1}^n$ and responses $\{y_i\}_{i=1}^n$ these models are estimated by solving\footnote{Not every model of interest falls exactly under problem \eqref{prob:pen_glm} e.g. the square root lasso loss function or combined penalties such as ElasticNet.} (or approximating)
\begin{equation} \label{prob:pen_glm}
\begin{aligned}
& \underset{\coef, \inter}{\textup{minimize}}  & & \frac{1}{n} \sum_{i=1}^n \loss(\coef^T x_i + \inter, y_i)  + \pen_{\tuneparam}(\coef),
\end{aligned}
\end{equation}
where $\loss$ is the loss function (linear, poisson, quantile regression, etc) and $\pen_{\tuneparam}$ is a penalty (lasso, group lasso, the SCAD function, etc).
A number of existing high quality GLM software packages are widely used such as  \Pglmnet \citep{regularization2010friedman} in R and \Psklearn in Python \citep{scikit2011pedrogosa}.
These frameworks, however, have not kept up with the last 20 years of statistics, machine learning, and optimization research.
We aim to bring these valuable developments from the text of papers and books to the fingertips of data analysts and researchers.

Beyond the basic ridge \citep{hoerl1970ridge} and lasso \citep{tibshirani1996regression} penalties, the last two decades have seen an abundance of new penalized regression methods.  
This includes \textit{structured sparsity} penalties such as group lasso \citep{yuan2006model}, fused lasso \citep{tibshirani2005sparsity},
generalized lasso \citep{tibshirani2011solution} nuclear norm \citep{negahban2011estimation, chen2013reduced}, and generalized ridge \citep{van2021lecture}. 
It also includes multiple penalties such as ElasticNet \citep{zou2005regularization, hu2018group}, sparse fused lasso \citep{tibshirani2005sparsity},  sparse group lasso \citep{simon2013sparse}, infimal overlapping group lasso \citep{jacob2009group}, and infimal row plus entrywise sparse \citep{tan2014learning}.

Significant progress has also been made on the computational front.
Fast optimization algorithms for GLMs include the \textit{fast iterative shrinkage-thresholding algorithm} (FISTA) \citep{beck2009fast}, \textit{alternating direction method of multipliers} (ADMM) \citep{boyd2011distributed}, advances in \textit{stochastic gradient descent} \citep{johnson2013accelerating}, and \textit{block proximal coordinate descent} \citep{tseng2001convergence, xu2017globally}. 

Convex penalties such as the lasso can return suboptimal parameter estimates. 
More accurate penalties have been developed with improved statistical properties including \textit{adaptive} \citep{zou2006adaptive} and non-convex penalties \citep{fan2001variable, zou2008one, breheny2011coordinate}. 
A major selling point of these penalties is that their increased accuracy often comes at limited additional computational expense \citep{fan2014strong, loh2017support}.
Furthermore, it is now understood that the right tuning parameter selection method (e.g. \textit{cross-validation} or an \textit{information criteria}) depends on whether the analyst's goal is prediction or model selection \citep{yang2005can, arlot2010survey, zhang2010regularization}.  
To use information criteria in practice we need estimates of the degrees of freedom \citep{tibshirani2012degrees, vaiter2017degrees} and in some cases the linear regression noise variance \citep{reid2016study}.

Most existing GLM software frameworks, however, have not been designed to comprehensively keep up with recent statistical and computational advances.
Much of modern methodology is scattered across different packages (see Appendix \ref{a:vs_existing}), but does not live under a unified framework.
For example, existing packages only support a limited number of loss + penalty combinations and cross-validation is often the only available tuning parameter selection method.
From a computational standpoint, there is no one ``best" optimization algorithm for everything; new methods and better software implementations are always being developed and not every loss + penalty combination can be fit by the same general optimization framework.\footnote{For example, the nuclear norm is non-separable so coordinate descent algorithms are not applicable.} 
Most GLM packages tie the user to one computational backend and do not allow users to use other optimization algorithms.

\subsection{Overview of the {\tt yaglm} package}

We develop \Pyaglm -- \textit{yet-another GLM package} -- to make the broader ecosystem of modern GLM methods accessible to data analysts: \url{https://github.com/yaglm/yaglm}.
Recent textbooks and references therein detail this ecosystem \citep{buhlmann2011statistics, hastie2019statistical, wainwright2019high, fan2020statistical}; Sections \ref{s:pen_glm_overview} and \ref{s:tuning} give an overview.
Existing software packages such as \Psklearn and \Pglmnet were not built with this broader ecosystem in mind.\footnote{\Pyaglm's development was inspired by and builds upon these existing packages!}

The \Pyaglm package supports fitting and tuning a large collection of loss + penalty and/or constraint combinations.
It provides structured sparsity, adaptive, and non-convex penalties as well as combinations of multiple penalties (see Section \ref{ss:multi_pen}).
\Pyaglm supports a number of tuning parameter selection methods including cross-validation and information criteria.
It includes estimators for degrees of freedom (Section \ref{ss:model_sel_and_dof}) and the linear regression noise variance (Section \ref{ss:lin_reg_noise}).

A handful optimization algorithms come built in including a flexible implementation of FISTA with adaptive restarts \citep{beck2009fast, o2015adaptive}, an augmented ADMM algorithm \citep{zhu2017augmented}, and a \Pcvxpy GLM solver \citep{diamond2016cvxpy}.
We also provide the \textit{local linear approximation} (LLA) algorithm for non-convex penalties that often comes with both strong computational and statistical guarantees \citep{zou2008one, fan2014strong}.
Our LLA implementation goes beyond the existing literature and generalizes to structured penalties (Section \ref{ss:lla}).
\textit{Path algorithms} as well as parallelization are supported for fast tuning.
\textbf{\Pyaglm is optimization algorithm agnostic; it is easy to employ any state of the art solver.}

The \Pyaglm package is designed to be user friendly.
It follows a unified \Psklearn compatible API \citep{api2013buitinck}, handles two stage estimators (Sections \ref{ss:adaptive} and \ref{ss:lla}), and automatically creates tuning parameter grids when possible.
It also supports sample weights, centering/scaling data standardization (even for sparse covariate matrices), and multiple response GLMs.

Beyond providing basic support for all the above features, \Pyaglm's modular design makes it easy to add new optimization algorithms, losses/penalties/constraints, and tuning parameter selection procedures.
New computational and statistical methods are always being invented/improved upon; \Pyaglm is a venue for researchers to make their contributions available to the broader data science community.

Section \ref{s:pen_glm_overview} gives an overview of penalized GLMs written for a reader familiar with the basics, but perhaps not with all the bells and whistles.
Section \ref{s:tuning} discusses tuning parameter selection. 
Section \ref{s:package} gives an overview of important software design choices. 
Appendix \ref{a:vs_existing} discusses comparisons with existing GLM software packages.
To our knowledge, no existing GLM package supports similar capabilities -- particularly for structured, adaptive and non-convex penalties.
Appendix \ref{a:pen_val_max} presents technical details related to automatic tuning parameter grid construction.

\subsection{Background and notation}
We loosely refer to the first term in problem \eqref{prob:pen_glm} as the GLM loss.
This includes standard GLM losses that are derived from \textit{exponential family} distributions (e.g. linear, logistic, poisson) as well as \textit{quasi-likelihoods} for M-estimators (e.g. quantile, cox, huber, SVM/hinge loss).
When the statistical model's likelihood plays an important role we will clarify that we are discussing an exponential family GLM.
For background on GLMs see the standard references cited above.

For a vector $x$, let $||x||_p$ denote the usual $L_p$ norm where $||x||_0$ counts the number of non-zero entries.
For a matrix $X$, let $\sigma(X)$ denote the vector of singular values where $\sigma_j(X)$ is the $j$th largest singular, let $\nuc{X} = \sum_{j} \sigma(X)_j$ denote the nuclear norm, and let let $ ||X||_{12} $ denote the usual $L1/L2$ mixed norm i.e. sum of euclidean norms of the rows.
$X(r, :)$ refers to the $r$th row of $X$ (similarly for $X(:, c)$).
Let $ \mathbf{1}_n \in \mathbb{R}^n_+$ be the vector of ones.
For two vectors, $a \le b$ denotes entrywise inequalities.
For a function $f: \mathbb{R} \to \mathbb{R}$ such as the absolute value function and vector $a \in \mathbb{R}^d$ we sometimes write $f(a) \in \mathbb{R}^d$ for the function applied entrywise to $a$.
The support of a vector $x$ is denoted $\support(x)$.

For the loss functions in Section \ref{s:glm_loss_and_pen_examples} we make the following standard definitions.
The sigmoid function $\sigmoid(z) := \frac{1}{1 - e^{-z}}$.
The softmax function $\soft: \mathbb{R}^{K} \to [0, 1]^K$,  $\soft(z)_k = \frac{e^{z_k} }{\sum_{j=1}^K e^{z_j}}$.
The huber function is given by
$$\huber(r) = 
\begin{cases}
\frac{1}{2} r^2 & \text{if } r \le \knot\\
\knot ( |r| - \knot) &  \text{if }  r > \knot,
\end{cases}
$$
where $\knot \ge 0$.
The tilted absolute value function is given by
$$\til(r; q) = 
\begin{cases}
q & \text{if } r \le 0\\
1 - q &  \text{if }  r > 0,
\end{cases}
$$
where $q \in [0, 1]$.

The proximal operator \citep{parikh2014proximal} of a function $g: \mathbb{R}^d \to \mathbb{R}$ with step size $\lr$ is
\begin{equation} \label{eq:proximal}
\prox_{g}(x; \lr) := \underset{z}{\textup{argmin}}  \;\; \frac{1}{2 \lr} ||x - z||_2^2 + g(z).
\end{equation}
We say informally that a function is \textit{proximable} if its proximal operator is easy to compute (e.g. see Chapter 6 of \cite{beck2017first}).
Computationally tractable proximal operators are a key ingredient for many optimization algorithms used with non-smooth objective functions.
For the definitions of standard optimization terms such as \textit{stationary points} or \textit{subgradient}, we point readers to standard optimization references such as \citep{beck2017first}.

The optimization problem \eqref{prob:pen_glm} needs to be solved repeatedly for a sequence of $\tuneparam$ values for parameter tuning.
A path algorithm uses warm starting to quickly solve the entire tuning sequence; it solves the problem for the first value, then uses that solution to warm start the algorithm for the next value and so forth \citep{friedman2007pathwise}.

\section{Penalized GLMs and other M-estimators}  \label{s:pen_glm_overview}

This section considers penalized problems (mostly) in the form of problem \eqref{prob:pen_glm} with covariates $x_i \in \mathbb{R}^d$ and responses $y_i \in \mathbb{R}$ ($y_i \in \mathbb{R}^{K}$ for multiple response loss functions), for observations $i=1, \dots, n$.
Here $\coef \in \mathbb{R}^d$ is a vector for single response outputs and a matrix $\coef \in \mathbb{R}^{d \times K}$ for multiple responses; similarly $\inter \in \mathbb{R}$ or  $\inter \in \mathbb{R}^K$.


We will denote the first term in the objective of problem \eqref{prob:pen_glm} as  $L(\coef, \inter)$.

\subsection{Common losses, penalties and constraints} \label{s:glm_loss_and_pen_examples}

Several common loss functions are shown in Table \ref{tab:loss}.
Some loss functions are not separable and cannot be written as in \eqref{prob:pen_glm}. 
For example, the loss from the \textit{square root lasso} is given by $L(\coef, \inter) = \frac{1}{\sqrt{n}}||y - X \coef + \mathbf{1}_n \inter ||_2 $  \citep{belloni2011square}.
For cox regression the loss function is $L(\coef, \inter)  = \frac{1}{n}\sum_{i \text{ s.t. } E_i=1} \left( x_i^T \coef + \inter - \log\sum_{j \text{ s.t. } y_j \ge y_i} e^{x_j^T \coef + \inter} \right) $ where $y_i \in \mathbb{R}_+$ is the $i$th subject's  survival time and $E_i \in \{1, 0\}$ indicates if $y_i$ is a death or censoring time, $E_i=1$ indicating the former (see Section 3.5 of \cite{hastie2019statistical}).


\begin{table}[H]
\centering
\begin{tabular}{|l|l|l|}
\hline
              & $\ell(z, y)$                                                                          & $y$ space                                                                              \\ \hline
Least squares & $\frac{1}{2}(y - z)^2$                                                                & $ y \in \mathbb{R}$                                                                    \\ \hline
Logistic      & $- y \log(\sigmoid(z)) + (1 - y) \log(1 - \sigmoid(z))$                               & $ y \in \{0, 1\}$                                                                      \\ \hline
Multinomial   & $\log(\soft(z)_{y})$                                                                  & $y \in \{1, \dots, K\}$                                                                \\ \hline
Poisson       & $e^z - z y$                                                                           & $ y \in \mathbb{R}_+$                                                                  \\ \hline
Huber         & $\huber_{\knot}(y - z)$                                                               & $ y \in \mathbb{R}$                                                                    \\ \hline
Quantile      & $\til_{q}(y - z)$                                                                     & $ y \in \mathbb{R}$                                                                    \\ \hline
Hinge         & $\text{max}(0, 1 - y z)$                                                              & $ y \in \{-1, 1\}$                                                                     \\ \hline
\end{tabular}
\caption{Several standard loss functions for problem \eqref{prob:pen_glm}; here $z = x^T \coef + \inter$.
The first four losses are exponential family GLM losses; the latter three are quasi-likelihoods.
The hinge loss is sometimes replaced by its smooth cousin, the squared hinge \citep{hastie2019statistical}; the quantile loss can also be replaced by a smooth proxy \citep{tan2021high}.
 }
\label{tab:loss}
\end{table}

We sometimes have sample weights, $\sampWeight \in \mathbb{R}^n_+$ in which case the loss function becomes $L(\coef, \inter)= \frac{1}{n} \sum_{i=1}^n \sampWeight_i \ell(\coef^T x_i + \inter, y_i) $.
We also allow observed offsets $\offset \in \mathbb{R}^n$ e.g. see Section 3 of \citep{hastie2019statistical}; in this case the loss becomes  $L(\coef, \inter)= \frac{1}{n} \sum_{i=1}^n  \ell(\offset_i + \coef^T x_i + \inter, y_i) $.
The one dimensional losses are typically extended to the multiple response case by summing over the coordinates of the response vector; this places a conditional independence assumption on the responses.

A number of standard convex penalty functions are listed in Table \ref{tab:basic_pen}.
This list is by no means exhaustive, but should give a sense of the main ideas.
\begin{itemize}

\item Given a list of groups,  $\groups = \{I_1, \dots, I_G\}$, $I_g \subseteq [d]$, the group lasso encourages entire groups of variables to go to zero (or not) together \citep{yuan2006model}.
We generally assume the groups are non-overlapping -- Section \ref{ss:multi_pen} discusses the overlapping case.
Typically the group weights are set to the square root of the group sizes i.e. $w_g =\sqrt{|\groups(g)|}$.
It can be helpful to preprocess the $X$ matrix with within group orthonormalization \citep{simon2012standardization}.

\item The \textit{multi-task} lasso is a special case of the group lasso where the groups are the rows of a matrix coefficient; here a feature is removed from all responses for a multi-response GLM.

\item Given a list of groups, the \textit{exclusive lasso} shrinks all entries towards zero, but ensures at least one variable in each group is non-zero \citep{zhou2010exclusive, campbell2017within}.

\item The nuclear norm encourages a low rank structure by shrinking the singular values of a coefficient matrix towards zero \citep{negahban2011estimation}.

\item The  \textit{total-variation}-1 (TV-1) penalty encourages successive entries to be equal \citep{tibshirani2005sparsity}.
As a generalization, given a graph whose nodes are the coefficient entries the fused lasso/graph fused lasso encourage nodes that are connected by an edge to be equal  \citep{tibshirani2011solution, sharpnack2012sparsistency, li2020graph}.

\item The generalized lasso -- a generalization of the fused lasso -- encourages sparsity in some linear transform, $\gen$ of the coefficient \citep{tibshirani2011solution, ali2019generalized}. 
The \textit{generalized ridge}\footnote{This is sometimes known as \textit{Tikhonov} regularization \citep{pyglmnet2020mainak}. Special cases often come with their own names such as ``Laplacian smoothing". It is often written as a quadratic form $\frac{1}{2}\coef^T M^TM \coef$.} is similar, but of course does not return exact zeros \citep{smola2003kernels, hebiri2011smooth, van2021lecture, green2021minimax}.


\end{itemize}

Quite a lot is known about the statistical properties of most of these convex penalties, see \citep{wainwright2019high, fan2020statistical} and the references therein.

\begin{table}[H]
\centering
\begin{tabular}{|l|l|l|l|l|}
\hline
                  & $\pen(\coef)$                                                           & Shape  & Parameters                                                  & Smooth \\ \hline
Lasso             & $\sum_{j=1}^d w_j |\coef_j|$                                            & vector & $w \in \mathbb{R}^d_+$                                      & no     \\ \hline
Group lasso       & $ \sum_{g=1}^G w_g ||\coef_{\groups(g)}||_2$                            & vector & $w \in  \mathbb{R}^G_+$, $\groups$                          & no     \\ \hline
Multi-task lasso  & $ \sum_{j=1}^d w_j ||\coef(j, :)||_2$                                   & matrix & $w \in \mathbb{R}^d_+$                                      & no     \\ \hline
Exclusive lasso   & $ \sum_{g=1}^G \left( \sum_{j \in \groups(g)}w_j |\coef_{j}| \right)^2$ & vector & $w \in  \mathbb{R}^d_+$, $\groups$                          & no     \\ \hline
TV-1              & $ \sum_{j=1}^{d-1} w_j |\coef_{j+1} - \coef_j|$                         & vector & $w \in  \mathbb{R}^{d-1}_+$                                 & no     \\ \hline
Graph fused lasso & $\sum_{(u, v) \in \text{edge list}} w_{uv} |\coef_{u} - \coef_v|$       & vector & $w \in  \mathbb{R}^{\text{num edges}}_+$, edge list         & no     \\ \hline
Generalized lasso & $||\text{diag}(w)\gen \coef||_1$                                        & vector & $\gen \in \mathbb{R}^{p \times d}$,  $w \in \mathbb{R}^p_+$ & no     \\ \hline
Nuclear norm      & $ \sum_{j=1}^d w_j \sigma_j(\coef)$                                     & matrix & $w \in \mathbb{R}^{\min(d, K)}$, decreasing entries         & no     \\ \hline
Ridge             & $\frac{1}{2} \sum_{j=1}^d w_j \coef_j^2$                                & vector & $w \in \mathbb{R}^d_+$                                      & yes    \\ \hline
Generalized Ridge & $\frac{1}{2} ||\tik \coef||_2^2$                                        & vector & $\tik \in \mathbb{R}^{p \times d}$                          & yes    \\ \hline
\end{tabular}
\caption{Basic convex penalties.
We include a weight term in the definition of these penalties due to its importance in later sections; the weights are set to 1 for the plain version of these penalties.
We indicate if the penalty naturally applies to a vector or matrix coefficient and if the penalty is differentiable (smooth).
}
\label{tab:basic_pen}
\end{table}

Several non-convex penalty functions are listed in Table \ref{tab:concave_pen}.
The lack of convexity introduces computational challenges since we are unable to obtain global minimizers in general. 
Section \ref{ss:lla} and \ref{ss:conconvex_direct} discuss computational strategies for handling non-convex penalties that often come with favorable statistical/computational guarantees.
These non-convex penalties can provide significantly more accurate estimates than their convex cousins (see references below).

\begin{table}[H]
\centering
\begin{tabular}{|l|l|l|}
\hline
               & $\pen_{\tuneparam}(\coef)$                                  & Shape  \\ \hline
Entrywise      & $\sum_{j=1}^d g_{\tuneparam}(|\coef_j|)$                    & vector \\ \hline
Group          & $ \sum_{g=1}^G  g_{\tuneparam}(||\coef_{\groups(g)}||_2)$   & vector \\ \hline
TV-1           & $\sum_{j=1}^{d-1} g_{\tuneparam} (|\coef_{j+1} - \coef_j|)$ & vector \\ \hline
Generalized    & $\sum_{r=1}^{p} g_{\tuneparam} (|\gen(r, :)^T \coef|)$    & vector \\ \hline
Multi-task     & $ \sum_{j=1}^d g_{\tuneparam}(||\coef(j, :)||_2)$           & matrix \\ \hline
Singular value & $ \sum_{j=1}^d g_{\tuneparam}( \sigma_j(\coef))$            & matrix \\ \hline
\end{tabular}
\caption{Examples of non-convex penalties.
For these penalties we must specify an increasing, concave function $g_{\tuneparam}: \mathbb{R}_+ \to \mathbb{R}$ such as the SCAD or MCP function \citep{fan2001variable, zhang2010nearly}.
These are sometimes called \textit{folded concave penalties} e.g.  since $g_{\tuneparam}(|x|)$ is not concave in $x$, but it is concave in $|x|$.
}
\label{tab:concave_pen}
\end{table}

We also consider constrained versions of  \eqref{prob:pen_glm} where the penalty function is replaced by a constraint.
Table \ref{tab:constraints} lists some common constraint sets used for GLMs \citep{xu2017generalized}.
These constraints have the appealing computational property that their projection operators are easy to compute, thus algorithms such as FISTA are straightforward to apply.

\begin{table}[H]
\centering
\begin{tabular}{|l|l|l|}
\hline
                & Set description                             & Convex \\ \hline
Positive        & $\coef \ge 0$                               & yes    \\ \hline
Box             & $a \le \coef \le b$                         & yes    \\ \hline
Simplex         & $\sum_{j=1}^d \coef_j = 1, \coef \ge 0$     & yes    \\ \hline
$L_q, q=1, 2$   & $||\coef||_q \le a$                         & yes    \\ \hline
Linear equality & $A \coef = b$                               & yes    \\ \hline
Isotonic        & $\coef_1 \le \coef_2 \le \dots \le \coef_d$ & yes    \\ \hline
Sparse          & $||\coef||_0 \le K$                         & no     \\ \hline
Rank            & $\text{rank}(\coef) \le K$                  & no     \\ \hline
\end{tabular}
\caption{Constraints with projection operators that are straightforward to implement.}
\label{tab:constraints}
\end{table}

\subsection{Multiple penalties: fully overlapping and infimal sums} \label{ss:multi_pen} 

It can be useful to combine multiple penalties.
For example, the ElasticNet penalty  \citep{zou2005regularization}, 
\begin{equation*}
 \tuneparam \cdot \enet \cdot ||\coef||_1  + \tuneparam \cdot (1 - \enet)  \cdot \frac{1}{2} ||\coef||_2^2
\end{equation*}
handles correlated features more gracefully than the lasso penalty alone, see Section 4.2 of \citep{hastie2019statistical}.
For example, if a group of features are highly correlated the lasso tends to arbitrarily pick one of these features from the group, while the ElasticNet tends to pick (or not) the entire group.
Here the mixing parameter $\enet \in [0, 1]$ controls the tradeoff between lasso and ridge penalties; it can either be set a-priori or tuned over a coarse\footnote{The ElasticNet parameterization has the computational advantage of having a smaller tuning parameter grid; instead of tuning over the product of two fine grids $\tuneparam_1 \times \tuneparam_2$, we can tune over one fine and one coarse grid $\tuneparam \times \enet$.}  grid e.g. $\{0.1, 0.2, \dots, 1\}$.
The group ElasticNet replaces the above entrywise lasso term with a group lasso term  \citep{hu2018group}. 

The sparse fused lasso \citep{tibshirani2005sparsity} both encourages successive features to be equal in addition to feature sparsity
\begin{equation*}
\tuneparam_1 ||\coef||_1 + \tuneparam_2 \sum_{j=2} |\coef_{j} - \coef_{j-1}|.
\end{equation*}
The sparse group lasso \citep{simon2013sparse} encourages both group and entrywise sparsity
\begin{equation*}
\tuneparam \cdot \enet ||\coef||_1  + \tuneparam \cdot  (1 - \enet) \cdot \sum_{g=1}^G w_g ||\coef_{\groups(g)}||_2,
\end{equation*}
and traditionally has the same parameterization as ElasticNet.
These are all examples of \textit{fully overlapping} penalties that take the form
\begin{equation*}
\pen_{\tuneparam_1, \dots, \tuneparam_q}(\coef) = \sum_{j=1}^q \pen^j_{\tuneparam_j}(\coef).
\end{equation*}


Another class of multiple penalties comes in the following form
\begin{equation} \label{prob:pen_glm_inf_sum}
\begin{aligned}
& \underset{b_1, \dots, b_q}{\textup{minimize}}  & &L\left(\sum_{j=1}^q b_j \right) +\sum_{j=1}^q \pen^j_{\tuneparam_j}(b_{j})
\end{aligned}
\end{equation}
where we have used $L(\cdot)$ to denote the loss function from problem \eqref{prob:pen_glm} as a function of $\coef$ and have dropped the intercept for the sake of exposition.
Here we have decomposed $\coef = \sum_{j=1}^q b_j$ and placed a different penalty on each term in the sum.

We refer to this as an  \textit{infimal sum}\footnote{This is our terminology -- we are not aware of an existing standard name for this class of penalties.} penalty since it is equivalent to the original problem \eqref{prob:pen_glm} with an infimal penalty,
\begin{equation*}
\pen_{\tuneparam_1, \dots, \tuneparam_q}(\coef) = \inf_{ \{b_j\}_{j=1}^q \text{ s.t. } \coef = \sum_{j=1}^q b_j} \left(\sum_{j=1}^q \pen^j_{\tuneparam_j}(b_{j}) \right).
\end{equation*}
Prominent examples of this kind of penalty for matrix shaped parameters include low rank plus sparse \citep{candes2011robust},
$$
\tuneparam_1 ||b_1||_*  + \tuneparam_2 ||b_2||_1,
$$
and row plus entrywise sparse \citep{tan2014learning}, 
$$
 \tuneparam_1 ||b_1||_{12} + \tuneparam_2 ||b_2||_1,
$$
where $\coef = b_1 + b_2$ in both cases.
When using the group lasso with overlapping groups the following infimal formulation is preferred over the naive application of the group lasso
$$
\pen_{\tuneparam}(\coef) = \inf_{\substack{\{b_g\}_{g=1}^G \text{ s.t. } \coef = \sum_{g=1}^g b_g, \\b_{g, \groups(g)^C} =0, g=1, \dots, G}} \left(\sum_{g=1}^G || b_{g, \groups(g)} ||_2 \right),
$$
see \cite{jacob2009group} or Section 4.3.3. of \cite{hastie2019statistical}.


\subsection{Adaptive penalties} \label{ss:adaptive}

A drawback of a convex penalty like the lasso is that it returns biased estimates (i.e. it shrinks the true non-zero coefficients towards zero) and can require restrictive assumptions to obtain good variable selection properties \citep{zou2006adaptive}.
The bias comes from the fact the lasso puts the same amount of penalty on both the true zeros as well as the true non-zero terms.
A  remedy for this issue is to use a weighted lasso penalty with intelligently chosen weights. 
The adaptive lasso implements this idea with a data driven choice for the weights \citep{zou2006adaptive} -- this idea also extends to other penalties \citep{wang2008note, chen2013reduced, viallon2013adaptive, ciuperca2017adaptive}.

We can rewrite problem \eqref{prob:pen_glm} with a weighted version of one of the non-smooth, convex penalties\footnote{The exclusive lasso does not fall under this framework, but this idea can be modified appropriately.} from Table \ref{tab:basic_pen} as
\begin{equation} \label{prob:weighted_lasso_prob}
\begin{aligned}
& \underset{\coef, \inter}{\textup{ minimize}}  & & \frac{1}{n} \sum_{i=1}^n \loss(\coef^T x_i + \inter, y_i)  + \tuneparam\sum_{j=1}^D  w_j |\trans(\coef)|,
\end{aligned}
\end{equation}
where $\trans: \mathbb{R}^d \to \mathbb{R}^D$ (or $\trans: \mathbb{R}^{d \times K} \to \mathbb{R}^D$ for matrix coefficients) is a transformation function, e.g.
\begin{itemize}
\item Entrywise penalty, $D=d$, $\trans(\coef)_j = \coef_j$

\item Group penalty, $D = G$, $\trans(\coef)_g = ||\coef_{\groups(g)}||_2$ 

\item Singular value penalty, $D = \min(d, K)$, $\trans(\coef) = \sigma(\coef)$.

\item Generalized lasso with matrix $\gen \in \mathbb{R}^{p \times d}$, $D = p$, $\trans(\coef) = \gen \coef$.
\end{itemize}

Adaptive penalties are two stage estimators. 
In the first stage, an initial estimate $\initCoef$ is obtained e.g. the lasso solution tuned via cross-validation.
The \textit{adaptive weights}  $w^{\text{adpt}} \in \mathbb{R}^{D}_+$ are then chosen as\footnote{The $\frac{1}{n}$ term takes care of division by zero issues coming from sparse initializers.} \citep{zou2009adaptive}
\begin{equation*}
w^{\text{adpt}} = \left(\trans(\initCoef) \right)^{-\adptExp} \text{ or } \left(\trans(\initCoef) + \frac{1}{n} \right)^{-\adptExp},
\end{equation*}
where $\adptExp > 0$ e.g. $0.5$ or $1$.
More generally, we can pick $w^{\text{adpt}}_j = \nabla g(s) \vline_{s = \trans(\initCoef)}$ where $g$ is an increasing, concave function. 
In the second stage we solve the weighted problem \eqref{prob:weighted_lasso_prob} using the adaptive weights. 
Pleasingly, we can often show that adaptive estimators satisfy the \textit{oracle property} -- meaning they perform as well as an unpenalized estimator fit to the true support \citep{zou2006adaptive, zou2009adaptive}.

A major selling point of adaptive penalties is that their strong statistical guarantees often come at the expense of only doubling the amount of computation needed for the standard penalty.
For example, we can use fast path algorithms to compute the tuning path of the adaptively weighted problem \eqref{prob:weighted_lasso_prob}. 
While problem \eqref{prob:weighted_lasso_prob} is of course convex when $t(\cdot)$ is convex, it is not convex for the singular value transform\footnote{Still, we can often still make use of proximal algorithms to solve problems with non-convex singular value penalties. Suppose $f(X) = \sum_{j=1} w_j \sigma_j(X)$ with $w_1 \le w_2 \dots$ or $f(X) = \sum_{j=1} g(\sigma_j(X))$ for some concave increasing function $g(\cdot)$. While $f$ is non-convex in both cases, we can often easily evaluate $f$'s proximal operator through the SVD of $X$ and the proximal operator of $g$. See \cite{chen2013reduced} and \cite{lu2015generalized, mazumder2020matrix} for details.} when $t(\cdot) = \sigma(\cdot)$ \citep{chen2013reduced}.

\subsection{Folded concave penalties and the LLA algorithm} \label{ss:lla}

This section considers penalized GLM problems with a non-convex penalty from Table \ref{tab:concave_pen} for some increasing, concave function $g_{\tuneparam}$.
As in the previous section we rewrite problem \eqref{prob:pen_glm} as 
\begin{equation} \label{prob:pen_glm_fcp}
\begin{aligned}
& \underset{\coef, \inter}{\textup{minimize}}  & & \frac{1}{n} \sum_{i=1}^n \ell(\coef^T x_i + \inter, y_i)  + \sum_{j=1}^D g_{\tuneparam}(|\trans(\coef)|).
\end{aligned}
\end{equation}
Following \citep{zou2008one, fan2014strong} we can use the two stage LLA algorithm to fit this penalty.

In the first stage  we compute an initial estimate $\initCoef$ e.g. the lasso solution tuned via cross-validation.
In the second stage we attempt to solve \eqref{prob:pen_glm_fcp} by running a \textit{majorization-minimization} (MM)  algorithm \citep{lange2000optimization, hunter2004tutorial} initialized from $\initCoef$.
In particular, given the current guess $b_{\text{current}}$ we obtain the \textit{surrogate function} (see references),
$$
Q(\coef | b_{\text{current}}) =  \sum_{j=1}^D \nabla g_{\tuneparam}(|\trans(b_{\text{current}})_j |) \cdot  |\trans(\coef)_j| + \text{constants},
$$
where $\nabla g_{\tuneparam}(\cdot)$ is the gradient.
The resulting algorithm is summarized in Algorithm \ref{algo:lla} below; following \citep{zou2008one, fan2014strong} we call this MM algorithm an LLA algorithm since we linearize the concave function.
After obtaining the surrogate function we solve a weighted problems of the form
\begin{equation} \label{prob:lla_weighted_sub_prob}
\begin{aligned}
& \underset{\coef, \inter}{\textup{minimize}}  & & \frac{1}{n} \sum_{i=1}^n \loss(\coef^T x_i + \inter, y_i)  + \sum_{j=1}^D w_j |\trans(\coef)|,
\end{aligned}
\end{equation}
where the \textit{majorization weights} are given by $w_j = \nabla g_{\tuneparam}(|\trans(b_{\text{current}})_j |)$.


\begin{algorithm}[H] \label{algo:lla}
\DontPrintSemicolon
 
 \KwInput{Data $\{ (y_i, x_i)\}_{i=1}^n$, tuning parameter $\tuneparam \ge 0$ and initializer $\estCoef{0}$}


\For{s=0, 1, 2, \dots}
{

Compute majorization weights, $w\spa{s} \in \mathbb{R}^D_+$
$$
w_j\spa{s} = \nabla g_{\tuneparam}(|\trans(\estCoef{s})_j|), j=1, \dots, D
$$

Update coefficient and intercept by solving the weighted subproblem
$$
\estCoef{s + 1}, \estInter{s+1}  = \text{Solution to problem \eqref{prob:lla_weighted_sub_prob} with weight vector } w=w\spa{s}
$$


}
\caption{The LLA algorithm for non-convex penalized problem \eqref{prob:pen_glm_fcp}}
\end{algorithm}

MM algorithms like Algorithm \ref{algo:lla} are only guaranteed to find a stationary point as opposed to a local/global minimizer \citep{lange2016mm}.
\cite{fan2014strong}, however, show the output of after \textbf{one step} of the algorithm is a stationary point of  \eqref{prob:pen_glm_fcp} with high-probability under broad conditions.
Moreover, \citep{zou2008one, fan2014strong} show that the output after \textbf{one step} can have strong statistical guarantees.
Both guarantees depend on having a ``good enough" initializer such as a lasso fit\footnote{We default to tuning with cross-validation, but the theoretical results of \citep{zou2008one, fan2014strong} do not take parameter selection into account.} and an appropriate penalty function like SCAD or MCP.
In some cases it is possible to initialize from 0 and obtain the same guarantees after two LLA steps.\footnote{If we initialize from zero then $\estCoef{1}$ is the $\tuneparam$-lasso fit when $g_{\tuneparam}$ is a SCAD-like function.}
Note \cite{zou2008one} and \cite{fan2014strong} only study\footnote{We expect their results to apply more generally, but we leave this to future work.} the entrywise case when $\trans(\cdot)$ is the identity.

%
%

\begin{remark}
Both the adaptive lasso in the previous section and LLA algorithm in this section require the same subroutine: an algorithm that solves the weighted problems of the form \eqref{prob:weighted_lasso_prob}/\eqref{prob:lla_weighted_sub_prob}.
\end{remark}

\subsection{Direct solution with non-convex penalties}\label{ss:conconvex_direct}

A number of papers propose directly solving problem \eqref{prob:pen_glm} when $\pen_{\tuneparam}$ is non-convex by running algorithms such as proximal coordinate descent \citep{mazumder2011sparsenet, breheny2011coordinate, zhao2018pathwise} or proximal gradient descent \citep{loh2015regularized}. 
While global solutions to concave penalized problem \eqref{prob:pen_glm} can have favorable statistical guarantees \citep{zhang2012general}, there is no guarantee we can ever find them.
This issue calls into question the relevance of statistical theory based on global solutions.

Recent statistical theory has begun to provide guarantees for the solutions we actually find in practice \citep{zhao2018pathwise}.
Remarkably, it is possible that \textit{every} stationary point of the problem has favorable statistical properties \citep{loh2015regularized, loh2017support}.
In contrast to the adaptive lasso and LLA algorithm, these guarantees do not require a good initializer; this has the computational benefit of not requiring a two stage algorithm.
Verifying that this theory applies requires a careful problem specific analysis.



\section{Tuning parameter selection} \label{s:tuning}

Fitting a penalized GLM \eqref{prob:pen_glm} of course requires selecting a value for the tuning parameter $\tuneparam$. 
For surveys of tuning methods see \citep{wu2020survey, fan2020statistical}.
Cross-validation and variants \citep{arlot2010survey, yu2014modified, chetverikov2021cross} are the most popular tuning methods, likely due to their simplicity and broad applicability.
Other tuning methods include: boostrap \citep{hall2009bootstrap}, generalized cross-validation (GCV) \citep{golub1979generalized}, risk minimization \citep{efron2004least, eldar2008generalized}, and information criteria (IC) such as AIC, BIC,  EBIC, HBIC \citep{chen2008extended, zhang2010regularization, fan2013tuning, wang2013calibrating}.
There are a handful of penalized methods were an a-priori known tuning parameter can be used such as the square root lasso \citep{belloni2011square}.

While IC and GCV methods have a computational advantage over CV, it's reasonable to wonder: why bother with anything but CV?
Section \ref{ss:pred_vs_model_sel} answers this question.
Section \ref{ss:model_sel_and_dof} discusses degrees of freedom estimates used by various model selection criteria.
Section \ref{ss:lin_reg_noise} discusses estimating the linear regression noise variance.

\subsection{Prediction vs. model selection} \label{ss:pred_vs_model_sel}

Two potential modeling goals include 1) \textit{prediction} and 2) \textit{model selection}.\footnote{More formally, \textit{loss efficiency} and \textit{model selection consistency} \citep{zhang2010regularization}.}
Suppose our observed data were generated by some true model coefficient $\coef^*$ and let $\widehat{\coef}_{\tuneparam}$ denote the estimate returned by solving \eqref{prob:pen_glm} with tuning parameter $\tuneparam$.
Good predictive performance of course means roughly $x^T \widehat{\coef}_{\tuneparam} \approx  x^T \coef^*$ for unseen values of $x$.
Good model selection performance means $\support(\widehat{\coef}_{\tuneparam}) \approx \support(\coef^*)$ i.e. identifying the true support of $\coef^*$ when the support is assumed to be sparse.\footnote{For structured sparsity such as the fused lasso model selection means correctly identifying the support of some other type of sparse structure.}
Unfortunately there is tension between these two goals; a tuning method cannot simultaneously be optimal for both prediction and model selection \citep{yang2005can}.

Methods like CV, GCV and AIC-like criteria tend to work best for prediction, but not model selection \citep{shao1993linear, wang2007tuning, zhang2010regularization, yu2014modified}.
For example, CV tends to select too many features since it is based on predictive accuracy \citep{wang2007tuning}.\footnote{It's better to under-shrink the true zeros -- these should be small anyways -- than over-shrink the true non-zeros.}
The ``1se" rule  for CV model selection --select the most shrunk model whose CV error is within one standard error of the best CV error -- is an attempt to address this issue \citep{friedman2009elements}.

Fortunately, BIC-like information criteria can have better model selection properties \citep{chen2008extended, wang2009shrinkage, zhang2010regularization, chen2012extended, kim2012consistent, fan2013tuning, wang2013calibrating, lee2014model, hui2015tuning}.
The performance results of BIC-like criteria critically depend on the assumption that the true model is in the solution path\footnote{In other words there exists some $\tuneparam$ such that $\support(\widehat{\coef}_{\tuneparam}) = \support(\coef^*)$.}; if this assumption fails then AIC-like information criteria may be more appropriate.

\subsection{Model selection criteria and degrees of freedom} \label{ss:model_sel_and_dof}

This section discusses model selection criteria for the broad set GLMs and M-estimators loosely in the form of \eqref{prob:pen_glm}.
We keep our discussion at a high-level and point to references for precise details.  

Many information and risk minimization criteria can be written as
\begin{equation} \label{eq:ic}
\text{measure of model fit}(\widehat{\coef}_{\tuneparam}) + (\text{tradeoff-factor}(n, d)) \cdot \text{measure of model complexity}(\widehat{\coef}_{\tuneparam}).
\end{equation}
Optimizing these criteria means finding the right balance between model fit and model complexity.
The tradeoff factor -- which we emphasize may be a function of both $n$ and $d$ -- controls this tradeoff.
For AIC-like criteria $\text{tradeoff-factor}(n, d)=2$, while for BIC-like criteria $\text{tradeoff-factor}(n, d) = \log(n)$.
The latter penalizes complex models more thus will select sparser coefficients.

Estimating (sometimes even defining) the measure of model complexity -- often called the \textit{effective degrees of freedom} DoF --takes some care.
Suppose we have an estimate $\widehat{\coef}_{\tuneparam}$ obtained by solving \eqref{prob:pen_glm} with some penalty.
For a sparsity inducing penalty such as the lasso a natural DoF estimate is $|\support(\widehat{\coef}_{\tuneparam})|$, the number of non-zero entries in the estimated coefficient.
Whether or not this is a good idea depends on the model under consideration; for example it can be a poor DOF estimator\footnote{Nevertheless, \cite{zhu2020polynomial} prove model selection consistence for an information criteria that uses the support size for best subset selection.} for best subset selection \citep{tibshirani2015degrees}.
For a formal definition of the DoF, {\tt DF}$_{\widehat{\coef}_{\tuneparam}}$,  see \citep{janson2015effective, tibshirani2015degrees, vaiter2017degrees} and references therein.
Note that number of the above references prove results for criteria that use the support size without any reference to this theoretical DoF quantity.


Consider the standard Gaussian linear regression setup where we observe $n$ iid observations, $y_i = x_i^T \coef^* +  \sigma \epsilon_i$, $\epsilon_i \sim N(0,1)$, $i=1, \dots, n$. 
We ignore the intercept for the sake of exposition.
For a ridge penalty there is a closed formula for a DoF estimate e.g. see Section 1.6 of \citep{van2021lecture}.
For a lasso penalty the support of the estimated coefficient is a good DoF estimate.
In particular, we have the remarkable result that $\mathbb{E}|\support(\widehat{\coef}_{\tuneparam})| = {\tt DF}_{\coef_{\tuneparam}}$ under mild conditions \citep{efron2004least, zou2007degrees, tibshirani2012degrees}. 
Analogous formulas hold for linear regression with a generalized lasso or ElasticNet penalty \citep{tibshirani2012degrees}. 
For an entrywise non-convex penalty (e.g. SCAD) a DoF estimate\footnote{We are not aware of formal results demonstrating directly this formula is a good DoF estimator.}is obtained via a ridge regression approximation \citep{fan2001variable}. 
In the large $n$ regime, \cite{zhang2010regularization} show this formula is well approximated by $|\support(\widehat{\coef}_{\tuneparam})|$.
Unbiased DoF estimates are available for other settings including: the  non-overlapping group lasso \citep{vaiter2012degrees}, convex constraints \citep{kato2009degrees}, a the nuclear norm/non-convex nuclear norm \citep{mazumder2020computing}.

Estimating the DoFs -- and developing model selection criteria -- for more general models takes some work for settings including: other exponential family GLMs, supervised M-estimators, structured sparsity penalties, and non-convex penalties.
\cite{park2007l1} gives a heuristic justification for using the support size for GLMs with a lasso penalty.
\cite{hui2015tuning} proposes the ERIC for the adaptive lasso with an exponential family GLM.
\cite{lee2014model} develops a BIC for lasso penalized quantile regression.
\cite{vaiter2017degrees} provides formulas for unbiased DoF estimations for exponential family distributions for the lasso, generalized lasso, group lasso, and generalized group lasso.
Several papers prove favorable estimation properties of certain ICs for non-convex entrywise penalties where the formulas implicitly use the size of the coefficient's support for a DoF estimate \citep{zhang2010regularization, fan2013tuning, wang2013calibrating}.
Another recent line of work develops develops a principled generalization of the AIC for exponential family GLMs with a lasso, entrywise non-convex penalty, and group lasso \citep{ninomiya2016aic, umezu2016consistency, umezu2019aic, komatsu2019aic}.

\subsection{Estimating the linear regression noise variance}\label{ss:lin_reg_noise}

Consider the standard Gaussian linear regression setup mentioned in the previous section and suppose we want to estimate the noise variance $\sigma^2$.
This quantity plays an important role in several high-dimensional inference procedures, see the references in Section 1 of \citep{yu2019estimating}.
Some model selection criteria need noise variance estimates\footnote{\cite{kim2012consistent} show that in some cases even a bad noise variance estimate will still lead to a consistent model selection criteria.} \citep{chen2008extended, kim2012consistent, hui2015tuning, ninomiya2016aic}, while others do not \citep{golub1979generalized, wang2007tuning, wang2013calibrating}.

It is straightforward to estimate $\sigma^2$ in the classical large $n$, small $d$ setting i.e. $\widehat{\sigma^2}_{\text{OLS}} := \frac{1}{n - d} ||y - X \widehat{\coef}_{\text{OLS}}||_2^2$ is an unbiased estimate where $\widehat{\coef}_{\text{OLS}}$ is the OLS estimate of the full model \citep{friedman2009elements}. 
In high-dimensional settings, however, this quantity is more challenging to estimate e.g. we typically have $\widehat{\sigma^2}_{\text{OLS}}=0$ when $d>n$. 
We mention several recently studied noise variance estimation methods that use an entrywise sparse penalty such as a lasso, adaptive lasso, or entrywise non-convex penalty.
\begin{enumerate}

\item Suppose we have an estimated coefficient $\widehat{\coef}_{\widehat{\tuneparam}}$ obtained by tuning the penalty (e.g. via cross-validation).
\cite{reid2016study} suggest using
$$\widehat{\sigma_{\widehat{\tuneparam}}^2} = \frac{1}{n - |\support( \widehat{\coef}_{\widehat{\tuneparam}})|} ||y - X \widehat{\coef}_{\widehat{\tuneparam}}||_2^2.$$

\item \cite{yu2019estimating} proposes noise variance estimators based on the \textit{natural lasso} and the \textit{organic lasso}. 
The former requires tuning a lasso (e.g. via cross-validation) while the latter can be fit with an a-priori known tuning parameter. 

\item  \cite{liu2020estimation} develop an estimate based on ridge regression with an a-priori selected tuning parameter.

\end{enumerate}
The ridge regression and organic lasso methods have the computational advantage of only requiring a single model fit  i.e. no parameter tuning needed.
Theoretical results, empirical comparisons, and pointers to other noise variance estimators may be found in the aforementioned papers and Section 3.7 of \cite{fan2020statistical}.
A few papers suggest simply plugging in $\widehat{\sigma_{\tuneparam}^2} = \frac{1}{n} ||y - X \widehat{\coef}_{\tuneparam}||_2^2$ when evaluating IC criteria \citep{hui2015tuning}.
In other words we use a different variance estimate at each value of $\tuneparam$.
Extensions of the above noise variance estimation methods to other structured settings such as generalized lasso and nuclear norm have received little attention to the best of our knowledge. 

\section{Package overview and design choices} \label{s:package}

The \Pyaglm package follows a unified \Psklearn compatible API \citep{api2013buitinck}.
To facilitate flexibility the \Pyaglm package is designed to be highly modular.
This section gives an overview of important design choices and assumes the reader is familiar with \Psklearn.
In \Psklearn each estimator is represented by an object; the estimator's parameters (e.g. the value of $\tuneparam$) are specified in the object's \coding{init} statement and the model is fit by calling the object's $\coding{.fit(X, y)}$ method.

The main estimator objects in \Pyaglm are \coding{Glm}, \coding{GlmCV},  and \coding{GlmCriteria}.
The \coding{Glm} class is reminiscent of \href{https://scikit-learn.org/stable/modules/generated/sklearn.linear_model.Lasso.html#sklearn.linear_model.Lasso}{sklearn.linear\_model.Lasso}; it represents a single fit for any loss (e.g. linear, logistic, quantile etc) plus penalty (e.g. lasso, group lasso, generalized Lasso) and/or constraint (e.g. positive, isotonic) combination.
The \coding{GlmCV} class is reminiscent of \href{https://scikit-learn.org/stable/modules/generated/sklearn.linear_model.LassoCV.html#sklearn.linear_model.LassoCV}{sklearn.linear\_model.LassoCV}; it represents any loss for a penalty tuned via cross-validation.
The \coding{GlmCriteria} class is similar, but tunes with a model selection criteria such as BIC (see Section \ref{s:tuning}).


In \Pyaglm several of the important estimator parameters (the loss, penalty, constraint and solver) are specified with configuration objects\footnote{Many of these can be specified with a string for users who just want the default settings.} as opposed to strings/floats.
See Example \ref{code:example} below.
E.g. the huber loss and its knot value are specified via the \coding{Huber(knot=2)}  config and an adaptive lasso is specified via \coding{Lasso} and \coding{Adaptive} configs.
The config objects both specify and document model settings e.g. the lasso's documentation lives in the  \coding{Lasso} object.
The penalty config objects are organized around the type of sparsity structure they try to induce (see Section \ref{ss:adaptive}) e.g. the \coding{Lasso} induces entrywise sparsity, the \coding{GroupLasso} induces group sparsity, etc.
Most of the non-smooth penalties described in Section \ref{s:pen_glm_overview} come in three \textit{flavors}: convex, adaptive, and non-convex.
For example, the lasso, the adaptive lasso and a SCAD are three flavors of an entrywise sparse penalty.

The extensive use of config objects is a critical design choice underlying the flexibility and breadth of the package.
They enable a modular design, documentation of complex parameters with sensible defaults, and the use of meta-algorithms\footnote{E.g. the LLA algorithm's subproblems are penalized  GLM problems where the penalty is a weighted convex penalty derived from the original non-convex penalty.} such as the adaptive lasso and the LLA algorithm \ref{algo:lla}.

\begin{code}
\captionof{listing}{Tuning an adaptive lasso with cross-validation}
\label{code:example}
\begin{minted}{python}
from yaglm import GlmCV
from yaglm.loss import Huber  # loss config
from yaglm.penalty import Lasso  # penalty config
from yaglm.penalty.flavors import Adaptive  # penalty flavor config
from yaglm.solver import FISTA  # solver config

est_cv = GlmCV(loss=Huber(knot=2), 
	       penalty=Lasso(flavor=Adaptive(expon=1)),  # tune an adaptive lasso 
	       init_est='default',  #  initializer for two stage estimators
	       standardize=True, #  center/scale -- true by default
	       solver=FISTA(max_iter=100),  # specify solver parameters 
	       cv=5, 
	       cv_select_rule='1se' ,
	       cv_n_jobs=-1 # parallelization -- use all available nodes
	       ).fit(X, y)  # fit just like sklearn!

\end{minted}
\end{code}

To make things concrete we walk through the code in Example \ref{code:example} that tunes huber regression with an adaptive lasso via cross-validation.

\begin{enumerate}

\item An initial estimator is fit to the raw data to obtain the adaptive weights.
By default the initializer is a lasso tuned via cross-validation (i.e. the same code is run except the Lasso's \coding{flavor} argument is excluded).
This happens before cross-validation so non-convex penalties use the same initializer in each fold.

\item The raw $X$ data are standardized via mean centering and standard deviation scaling.
The same standardization is gracefully applied to sparse $X$ matrices through the\\ \href{https://docs.scipy.org/doc/scipy/reference/generated/scipy.sparse.linalg.LinearOperator.html}{scipy.sparse.LinearOperator} class \citep{sicpy2020virtanen}.

\item The tuning parameter grid is created automatically.
This requires computing the largest reasonable tuning parameter value (if available) for the given loss, penalty and processed data (see Appendix \ref{a:pen_val_max}).
The user can specify a custom tuning grid via the penalty config object's \coding{.tune()} method. 

\item The raw data are split into 5 cross-validation folds, preprocessed, and problem \eqref{prob:pen_glm} is solved for each tuning parameter value in the tuning grid using the specified FISTA solver.
A path algorithm is used to compute the tuning path quickly for each fold if the solver supports a path algorithm.
If a path algorithm is available the computation is parallelized over the folds; otherwise it is parallelized over all fold-tune parameter combinations.
Cross-validation test/training metrics are computed on the fitted coefficients.

\item The optimal tuning parameter is selected using the `1se' rule from the CV results.

\item Problem \eqref{prob:pen_glm} is solved with the processed data for the optimal tuning parameters using the specified solver.

\end{enumerate}

The backend steps for \coding{GlmCriteria} are similar, except the selected model does not need to be refit after tuning parameter selection since it was already computed during tuning.
Some model selection criteria require estimators for the degrees of freedom (Section \ref{ss:model_sel_and_dof}) and possibly for the linear regression noise variance (Section \ref{ss:lin_reg_noise}).
These estimates are obtained using an \coding{Inferencer} object that can be passed to \coding{GlmCriteria}'s \coding{init} statement.

%
%

The built in FISTA framework was inspired by the \Ppyunlocbox package \citep{defferrard2017pyunlocbox}.
It can solve GLM problems with a smooth loss and a proximable penalty (e.g. it does not cover the quantile loss or generalized lasso penalty).
Proximal penalties include: lasso, group lasso, exclusive lasso \citep{lin2019dual}, nuclear norm, ElasticNet, sparse group lasso, and any separable or infimal sum of proximable penalties.
Our FISTA implementation automatically computes step sizes for Lipschitz differentiable loss functions and does a backtracking line search for non-Lipschitz  differentiable losses (e.g. poisson).
We include the augmented ADMM algorithm of \citep{zhu2017augmented} for some generalized lasso problems and another computational backend based on  \Pcvxpy \citep{diamond2016cvxpy, agrawal2018rewriting}.
We also provide the LLA algorithm (see Section \ref{ss:lla}), which can solve most non-convex structured problems. 

The \Pyaglm package is optimization algorithm agnostic; its modular design makes it easy to use different computational backends.
We make it straightforward for users to provide their favorite solvers by simply wrapping them in a \coding{SolverConfig} object and passing that to the estimator classes.
This enables data analysts to have easy access to state of the art optimization algorithms.
It also means optimization experts can focus their efforts on algorithm development and let \Pyaglm take care of the rest.\footnote{\Psklearn's LassoCV class accomplishes this for the important case of linear regression with a lasso penalty.}

Wherever possible we provide sensible defaults so most users can obtain good performance with little cognitive overhead.
This includes default optimization solvers,  automatic tuning parameter grid creation (when possible), and initializers for two stage estimators.
Other parameters can be tuned over including loss parameters (e.g. the huber knot) and other penalty parameters (e.g.  SCAD's ``a" parameter).
Lots of customization is available e.g. custom evaluation metrics can be specified via the \coding{cv\_metrics} argument to \coding{GlmCV}.

We attempt to make the frontend and backend code as readable as possible e.g. we use interpretable and standardized names such as \coding{pen\_val}. 
Additional examples and documentation are available on the Github repository.

\section{Conclusion and future directions} \label{s:conclusion}

The \Pyaglm package updates the Python ecosystem with a comprehensive and modern penalized GLM package.
The main contribution of this package is a framework that brings everything together (e.g. combinations of losses + penalties, optimization algorithms, automatic tuning parameter grid creation, parameter tuning logistics, etc) and makes it usable by data analysts and researchers. 

At the time of this writing we have included several default optimization algorithms and tuning parameter selection methods.
Going forward we will continue to add state of the art solvers (e.g. \Pceler), tuning methods, and penalties \citep{carmichael2021folded}.
We are also in the process of improving the quality of the software with documentation, testing, and continuous integration tools.

In the medium/long term we aim to add: a block coordinate descent backend based on \citep{xu2017globally}, support for estimating mixed, exponential family graphical models via neighborhood selection \citep{yang2014mixed}, classical statistical inference for unpenalized GLMs (e.g. p-values), and modern statistical inference for penalized GLMs such as \textit{stability selection} \citep{meinshausen2010stability}.

\section*{Acknowledgements}
IC was supported by the National Science Foundation under Award No. 1902440.
We thank Mainak Jas for suggesting the use of config objects, which turned out to be a critical design feature for the package.

\appendix

\section{Comparisons with existing software packages} \label{a:vs_existing}

The \Pyaglm package supports features that typically go beyond existing packages including:
\begin{itemize}
\item Many combinations of loss + penalty and/or constraints
\item Adaptive and non-convex flavors of the convex penalties
\item Combinations of multiple penalties (see Section \ref{ss:multi_pen})
\item Automatic tuning parameter grid creation
\item A variety of tuning parameter selection methods
\item The tuning backend supports parallelization and path algorithms
\item Tuning diagnostics and visualizations
\item A flexible optimization backend tailored to GLMs/supervised M-estimators built on FISTA, ADMM and \Pcvxpy
\item The LLA algorithm for (possibly structured) non-convex penalties
\item Any user provided optimization algorithm
\end{itemize}

To the best of our knowledge the two most popular existing GLM packages are \Psklearn \citep{scikit2011pedrogosa} in Python and \Pglmnet \citep{regularization2010friedman} in R.
While \Pyaglm was designed to be as close to \Psklearn as possible, but goes well beyond \Psklearn's capabilities.
\Psklearn supports a number of the above features for linear regression (e.g. fast path algorithm, somewhat customizable solvers, a few convex penalties such as lasso, ElasticNet and multi-task lasso).
For losses beyond linear regression, \Psklearn supports a limited number of penalties (e.g. usually just ridge and sometimes lasso) and has no support for automatic tuning parameter grid creation.
There is no support for adaptive or non-convex penalties and almost no support for tuning parameter selection beyond cross-validation.
Except for linear regression, \Psklearn estimator objects were not designed to facilitate user supplied solvers


The \Pglmnet package supports several loss functions (e.g. logistic, poisson, etc) with ElasticNet penalties.
Its computational backend is a (famous) coordinate descent algorithm and tuning parameter selection is performed via cross-validation. 
There is no support for adaptive/non-convex penalties, tuning parameter selection beyond CV, or custom solvers.

There are several package that support non-convex penalties for several loss functions including \Ppicasso \citep{ge2019picasso}, \Pandy  \citep{massias2020dual}, \Pgrpreg \citep{breheny2015group}, and \Pncreg \citep{breheny2011coordinate}.
These packages still do not include the large number of losses and penalties supported in \Pyaglm and typically do not provide tuning parameter selection methods beyond CV.
The \PGGMncv package  \citep{williams2020ggmncv}  fits a gaussian graphical model with a non-convex penalty via the LLA algorithm and supports a variety of tuning parameter selection methods.
 \Plassopack \citep{ahrens2020lassopack} is a Stata package that supports adaptive lasso penalties for linear regression and includes several tuning parameter selection methods.

Other packages worth mentioning that provide fast solvers for several penalized GLM losses include  \Plightning \citep{lightning2016blondel}, \Pceler \citep{celer2018massias}, and \Ppyglmnet  \citep{pyglmnet2020mainak}.
The popular \Pstatsmodels package \citep{seabold2010statsmodels} supports a variety of unpenalized GLMs. 
The \Pgenlasso package has good support for generalized lasso problems with a linear regression loss \citep{arnold2020genlasso}.

Several of these packages have better optimization algorithms than the ones that come built in with \Pyaglm.
The sophistication of these solvers, however, tends to limit their breadth e.g. a given solver may not support non-separable penalties, losses beyond linear regression, or non-smooth losses.
However state of the art solvers can be used as a computation backend for \Pyaglm!

\section{Automatic tuning grid creation and specifying the largest tuning parameter value} \label{a:pen_val_max}

To make the \Pyaglm package user friendly we automatically create the tuning parameter grids whenever possible.
Users may also provide their own tuning grids.

For most penalty parameters, we tune  $\tuneparam$ along  a sequence of points in an interval $[\epsilon  \tuneparam_{\text{max}}, \tuneparam_{\text{max}}]$ for some small\footnote{By setting $\epsilon > 0$ we exclude the unpenalized case when $\tuneparam=0$ intentionally to avoid computational issues when $d > n$.} value of $\epsilon$ (e.g. $10^{-3}$).
By default we create a logarithmically spaced sequence of (e.g. 100) points in this interval (meaning the points are more dense around the smaller values).

Given dataset $(X, y)$, loss $\ell(\cdot)$ and penalty $\pen_{\lambda}(\cdot)$ we can often (but not always) cheaply compute an intelligent choice for the largest reasonable tuning value $\tuneparam_{\text{max}}$.
This rest of this section discusses automatic specification of the $\tuneparam_{\text{max}}$ value.
Section \ref{as:pen_val_max__lasso} discusses convex and non-convex shinkage penalties.
Section \ref{as:pen_val_max__lla} discusses non-convex penalties fit with the LLA algorithm.
Section \ref{as:pen_val_max__lin_reg_ridge} presents an approach for linear regression with a ridge penalty; Section \ref{as:pen_val_max__lin_reg_glm} extends this via a heuristic to other loss functions.

This section does not cover every loss + penalty combination e.g. we are not aware of a simple method for the generalized lasso.
Most of the results in this section are not novel \citep{park2007l1}, but we are not aware of another approachable, general description of these details.

\subsection{Non-smooth shrinkage penalties} \label{as:pen_val_max__lasso}

Suppose $\pen_{\tuneparam}(\cdot)$ non-smooth at 0 and consider the following minimization problem
\begin{equation*}
\begin{aligned}
& \underset{\coef, u}{\textup{minimize}}  & & L(\coef, u)  + \pen_{\tuneparam}(\coef),
\end{aligned}
\end{equation*}
where $\coef$ is penalized, but $u$ is not e.g. $u$ might be the intercept in \eqref{prob:pen_glm}.
For a non-smooth function $g: \mathbb{R}^d \to \mathbb{R}$ let $\nabla g(x) \subseteq \mathbb{R}^d$ denote the set of \textit{sub-gradients} (called the \textit{sub-differential}) e.g. see Chapter 3 of \citep{beck2017first}.
For example, $\nabla |0|= [-1, 1]$.

\begin{definition} \label{def:KLB}

We say $\tuneparam_{killer-lbd}$ is a {\normalfont weak killer lower bound (KLB)} if for all $\tuneparam \ge \tuneparam_{killer-lbd}$ there is a solution to 
\begin{equation*}
- \nabla_{\coef} L(0, u^*) \in  \nabla \pen_{\tuneparam}(0)
\end{equation*}
for some $u^* \in \mathbb{U}_0^* := \underset{u}{\textup{argmin}} \;\;  L(0, u)$.
We say $\tuneparam_{killer-lbd}$ is a  {\normalfont strong killer lower bound} if $0$ is the only solution i.e. if
\begin{equation} \label{eq:strong_klb_unique_set}
\{\coef | -\nabla_{\coef} L(\coef, u^*) \in \nabla \pen_{\tuneparam}(\coef), u^* \in \mathbb{U}_0^*  \} = \{0\}.
\end{equation}
\end{definition}

A strong KLB guarantees that $\lambda$ is sufficiently large such that $\coef=0$ at every stationary point.
A weak KLB only guarantees the existence of such a stationary point.
Of course if $L$ is strictly convex then any weak KLB is automatically a strong KLB and $\coef=0$ is the global minimizer.

Returning to the penalized GLM problem \eqref{prob:pen_glm} we can often obtain
\begin{equation} \label{eq:inter_sol_at_0}
\interSolAtZero :=  \underset{\inter}{\textup{argmin}} \;\; L(0, \inter)
\end{equation}
for standard GLM loss functions.
E.g. $\interSolAtZero = \frac{1}{n} \sum_{i=1}^n y_i$ for linear, logistic and poisson regression.

We can then use standard results about sub-differentials to obtain KLBs  \citep{beck2017first}.
For example, we can appeal directly to the sub-differential of the L1 norm for a lasso penalty to obtain
\begin{equation} 
 \nabla_{\coef} L(0, \interSolAtZero)  \in \tuneparam \nabla ||0||_1 \iff ||\nabla_{\coef} L(0, \interSolAtZero)||_{\text{max}} \le \tuneparam.
\end{equation}

When $\pen_{\tuneparam}$ has a nice proximal operator we can easily compute a KLB as
%
$$
\tuneparam_{killer-lbd} = \min \{ \tuneparam \; |  \; 0 \in \prox_{\pen_{\tuneparam}}\left( -L(0, u^*) \right) \}.
$$
The above lasso KLB results can be obtained easily from the fact the proximal operator of the L1 norm is the soft-thresholding operator $\prox_{\tuneparam ||\cdot||_1}(x) =   \text{sign}(x) \odot \max(|x| - \tuneparam, 0)$.

Table \ref{tab:lasso_killer_lbd} summarizes KLBs for several penalties with an optional strictly positive weight vector (recall Table \ref{tab:basic_pen}). 
If some of the weights are zero then these formulas are no longer exact.\footnote{Let $\mathcal{S} = \text{support}(w)$ and suppose we count the terms in $\mathcal{S}^C$ among the unpenalized variables in Definition \ref{def:KLB}.
Then one could obtain exact formulas by first solving $\underset{\inter, \coef}{\textup{argmin}} \;\; L(\coef, \inter) \text{ s.t. } \coef_{\mathcal{S}^C} = 0$ the deriving similar formulas as in Table \ref{tab:lasso_killer_lbd}.}

\begin{table}[H]
\centering
\begin{tabular}{|l|l|l}
\cline{1-2}
             & $\killerLbd$                                                                                 &  \\ \cline{1-2}
Lasso        & $|| \nabla_{\coef} L(0, \interSolAtZero) \odot \frac{1}{w} ||_{\text{max}}$                  &  \\ \cline{1-2}
Group lasso  & $\max_{g \in [G]} \; \frac{1}{w_g} ||\nabla_{\groups(g)} L(0, \interSolAtZero) ||_2$         &  \\ \cline{1-2}
Nuclear norm & $||\sigma\left(\nabla_{\coef} L(0, \interSolAtZero) \right)\odot \frac{1}{w}||_{\text{max}}$ &  \\ \cline{1-2}
\end{tabular}
\caption{Killer lower bounds for several non-smooth convex penalties with a strictly positive weight vector $w > 0$.
For the nuclear norm we assume the entries of $w$ are non-decreasing e.g. if they are adaptive weights (see Section \ref{ss:adaptive} and the footnote about the proximal operator).
For the entrywise lasso, group lasso, and unweighted nuclear norm these are strong KLBs when $n > d$ and $X$ has full rank; otherwise they are weak KLBs.
}
\label{tab:lasso_killer_lbd}
\end{table}


\begin{remark} \label{rem:adpt_lasso_killer_lbd}
For the adaptive lasso we can use the formulas from Table \ref{tab:lasso_killer_lbd} with the adaptive weights to obtain a killer lower bound. 
\end{remark}

Recall the SCAD and MCP both look like a lasso around the origin i.e. $\nabla\pen_{\tuneparam}(0) = [-\tuneparam, \tuneparam]$ in both cases.
The formulas in \eqref{tab:lasso_killer_lbd} therefore give weak KLBs for the SCAD/MCP versions of these penalties.

\subsection{The LLA algorithm}\label{as:pen_val_max__lla}

For non-convex penalties fit using the LLA algorithm we can obtain a stronger version of $\killerLbd$ than in the previous section.
Let $\initCoef$ be the initializer for the LLA algorithm.
The following proposition guarantees that if $\tuneparam \ge \killerLbd$ then the LLA algorithm initialized from $\initCoef$  converges to 0 in one step.

\begin{proposition} \label{prop:killer_lbd_lla}
Consider using the LLA algorithm for problem \eqref{prob:pen_glm_fcp} with a transformation $\trans(\cdot)$ such that $t(x) = x \iff x = 0$.
Suppose $\killerLbd$ is a strong KLB for the unweighted version of problem \eqref{prob:weighted_lasso_prob} and $\pen_{\tuneparam}$ is a SCAD-like penalty satisfying Definition 3.1 of \cite{carmichael2021folded} with parameters $(a_1, b_1)$.
Let
\begin{equation} \label{eq:killer_lbd_lla}
\killerLbd^{\text{LLA}} := \max \left( \frac{||\trans(\initCoef)||_{\text{max}}}{b_1},  \frac{\killerLbd}{a_1} \right).
\end{equation}
Then for all $\tuneparam \ge \killerLbd^{\text{LLA}}$, taking one step LLA step from $\initCoef$ will output zero and so will the next LLA step i.e. the algorithm has converged.

\end{proposition}

\subsection{Linear regression with ridge penalty} \label{as:pen_val_max__lin_reg_ridge}

For ridge regression we cannot obtain exact zeros, but we can force the estimated coefficient to be as small as we like.
In particular, for a given dataset $X, y$ and user specified $\epsilon$ we can compute a bound such that $\lambda \ge  \tuneparam_{\text{max}}  \implies ||\widehat{\coef}_{\tuneparam}||_2 \le \epsilon$.

We first consider the case with no intercept in the model.
Recall the closed form solution to linear regression with a ridge penalty is
\begin{equation} \label{eq:lin_reg_ridge_soln}
\begin{aligned}
\widehat{\coef}_{\tuneparam}  & = \underset{\coef}{\textup{argmin}}  \;\;  \frac{1}{n} \sum_{i=1}^n ( y_i - \coef^T x_i)^2  + \tuneparam \frac{1}{2} ||\coef||_2^2
 = (X^T X + n \tuneparam I_d)^{-1} X^T y
\end{aligned}
\end{equation}
Next let $X = U \text{diag}(\sigma) V^T$ be the SVD decomposition of $X$.
We can then check the following holds for any $K \le \text{min}(n, d)$,
\begin{equation} \label{eq:lin_reg_ridge_norm_formula}
\begin{aligned}
||\widehat{\coef}_{\tuneparam}||_2^2 
& = \sum_{k=1}^{\text{min}(n, d)} \left( \frac{\sigma_k}{\sigma_k^2 + n \tuneparam}  \cdot U_k^T y \right)^2 \\
& \le  \sum_{k=1}^{K} \left( \frac{\sigma_k}{\sigma_k^2 + n \tuneparam}  \cdot U_k^T y \right)^2 + (\text{min}(n, d) - K) \cdot \left( \frac{\sigma_K  ||y||_2}{n \lambda} \right)^2,
\end{aligned}
\end{equation}
by properties of the SVD and Cauchy-Schwarz.
While the first equation on the right of \eqref{eq:lin_reg_ridge_norm_formula} gives an exact expression, it requires the full SVD of $X$ which may be computationally expensive.
The second equation gives an upper bound that only depends on the leading user specified $K$ singular values thus is cheaper to compute.
The second bound may be loose if $K$ is too small as to ignore large singular values or  $y$ is strongly correlated with the lower principal components of $X$.
Both equations are smoothly decreasing in $\tuneparam$ so it is straightforward to numerically obtain $\tuneparam_{\text{max}}$.

We mention one more bound,
\begin{equation}\label{eq:ridge_coef_ubd}
||\widehat{\coef}_{\tuneparam}||_2 
%
\le \frac{||X^T y||_2}{ \sigma_{\text{min}}^2  + n \tuneparam},
\end{equation}
that follows by standard properties of the operator norm.
An advantage of this bound is that it only needs one singular value. 
Equation \eqref{eq:ridge_coef_ubd} can be analytically inverted to obtain\footnote{In the case this formula suggests $\tuneparam_{\text{max}}=0$ we default to $ \frac{1}{n}\sigma_{\text{min}}^2$.} $\tuneparam_{\text{max}}$.

When an intercept is included in the model we replace\footnote{When $\widehat{\coef}_{\tuneparam}$ we assume the intercept is approximately the mean of the observations.} $y$ above with $y - \overline{y}$ in \eqref{eq:lin_reg_ridge_norm_formula} and \eqref{eq:ridge_coef_ubd} above (if $||\coef||$ is small then $\inter \approx \overline{y}$).
The above formulas can be modified to handle sample weights, $s \in \mathbb{R}^n_+$, by replacing the $X, y$ data with the transformed
$$
\text{diag}(s)^{1/2}X \text{ and } \text{diag}(s)^{1/2} y.
$$

We can modify the above approach for a weighted ridge penalty $\sum_{j} w_j \coef_j^2$.
For a strictly positive penalty weight vector $w \in \mathbb{R}^d_+$ we can check\footnote{Making the substitution $\alpha = \sqrt{w} \odot \coef$ in the ridge problem \eqref{eq:lin_reg_ridge_soln}, we obtain an analytically solvable ridge regression problem in the transformed data.} the weighted ridge solution is given by
$$
\widehat{\coef}_{w, \tuneparam} = \text{diag}(w)^{-1/2} \widetilde{\coef}_{\tuneparam}
$$
where $ \widetilde{\coef}$ is the ridge solution \eqref{eq:lin_reg_ridge_soln} using the transformed data matrix $\widetilde{X} = X \text{diag}(w)^{-1/2}$.
We can then bound $||\widehat{\coef}_{w, \tuneparam}||_2$ using  \eqref{eq:lin_reg_ridge_norm_formula} by replacing the equality with an inequality and putting a $\frac{1}{||w||_{\text{min}}}$ on the right hand side.
Similarly for \eqref{eq:ridge_coef_ubd}.

\subsection{Smooth GLM loss with a ridge penalty, Newton step heuristic}\label{as:pen_val_max__lin_reg_glm}

In this section we present a heuristic to determine $\tuneparam_{\text{max}}$ for twice differentiable loss functions with a ridge penalty,
$$
 \frac{1}{n} \sum_{i=1}^n \ell(\coef^T x_i + \inter, y_i)  + \tuneparam \frac{1}{2} ||\coef||_2^2.
$$
Unfortunately we cannot repeat the argument in Section \ref{as:pen_val_max__lin_reg_ridge} because there is no closed formula for the ridge solution in general.
We instead develop a heuristic based on the Newton update.


Our heuristic is based on constructing \textbf{an upper bound for $||\coef\spa{1}_{\text{Newton}}||_2$ where $\coef\spa{1}_{\text{Newton}}$ is the result of taking a single Newton step in the coefficient from 0} holding fixed\footnote{If $\coef \approx 0$ then the intercept solution is approximately $\interSolAtZero$ defined in \eqref{eq:inter_sol_at_0}.} $\inter =  \interSolAtZero$.
We can check this Newton update step is given by
\begin{equation*}
\coef\spa{1}_{\text{Newton}} = \left(X^T \text{diag}(h(\interSolAtZero, y) ) X + n \tuneparam I_d  \right)^{-1} X^T g(\interSolAtZero, y)
\end{equation*}
where $g(\cdot, y)_i = \ell'(\cdot, y_i) $ and $h(\cdot)_i =  \ell''(\cdot, y_i) $ for $i=1, \dots, n$.
This is exactly the formula for the ridge linear regression solution \eqref{eq:lin_reg_ridge_soln} with the dataset
\begin{equation*} 
 \text{diag}(h(\interSolAtZero, y) )^{1/2} X \text{ and }   \text{diag}(h(\interSolAtZero, y) )^{-1/2} g(\interSolAtZero, y)
\end{equation*}
We can therefore use the bounds \eqref{eq:lin_reg_ridge_norm_formula} or \eqref{eq:ridge_coef_ubd} from the previous sections to obtain bounds on $||\coef\spa{1}_{\text{Newton}}||_2$.
These bounds can then be used to obtain a value $\tuneparam_{\text{max}}$ that enforces $||\coef\spa{1}_{\text{Newton}}||_2 \le \epsilon$.
\section{Proofs} \label{s:proofs}


\begin{proof} of Proposition \ref{prop:killer_lbd_lla}

By Definition 3.1 of \citep{carmichael2021folded}, $x \le b_1 \tuneparam \implies \nabla g_{\tuneparam}(x) \ge a_1 \tuneparam$ for any $x \ge 0$.
Therefore, $||t(\coef\spa{0})||_{\text{max}} \le b_1 \tuneparam \implies ||\nabla g_{\tuneparam}(|t(\coef\spa{0})|)||_{\text{min}} \ge a_1 \tuneparam$ for any $\tuneparam$.
Now suppose $\tuneparam \ge \killerLbd^{\text{LLA}}$.
By the construction of $\killerLbd^{\text{LLA}}$ in  \eqref{eq:killer_lbd_lla} we have
$
\frac{||t(\coef\spa{0})||_{\text{max}} }{b_1} \le  \killerLbd^{\text{LLA}} \le \tuneparam
\implies 
 ||\nabla g_{\tuneparam}(|t(\coef\spa{0})|)||_{\text{min}} \ge a_1 \tuneparam \ge a_1 \killerLbd^{\text{LLA}} \ge \killerLbd.
 $

Let $w\spa{0} = \nabla g_{\tuneparam}(|t(\coef\spa{0})|)$ be the majorization weights at $\initCoef$ in Algorithm \ref{algo:lla}.
By the above argument, if $\tuneparam \ge \killerLbd^{\text{LLA}}$ then $||w\spa{0}||_{\text{min}} \ge \killerLbd$.
It's not hard to see that by strong KLB assumption the global solution to problem \eqref{prob:lla_weighted_sub_prob} with $w=w\spa{0}$ is 0.
Therefore $\coef\spa{1} = 0$.

Since $g_{\tuneparam}(\cdot)$ is concave, $\nabla g_{\tuneparam}(|t(0)|) \ge \nabla g_{\tuneparam}(|t(\coef\spa{0})_j|)$ for any $j$.
The same argument as above then gives $\estCoef{2} = 0$.


%
%
\end{proof}

\bibliographystyle{apalike}
\bibliography{refs}

\end{document}